\documentclass[aps,prd,groupedaddress,amssymb,eqsecnum,showpacs,epsfig,nofootinbib]{revtex4}
\usepackage{graphicx}
\usepackage{bm}
\usepackage{dcolumn}
\usepackage{amsmath}

\usepackage{amsmath}
\usepackage{amssymb}

\numberwithin{equation}{section}
\usepackage{epsfig}

\def \lleq {\lower0.9ex\hbox{ $\buildrel < \over \sim$} ~}
\def \ggeq {\lower0.9ex\hbox{ $\buildrel > \over \sim$} ~}

\def \omm  {\Omega_{0 {\rm m}}}

\def \beq  {\begin{equation}}
\def \eeq  {\end{equation}}
\def \ber  {\begin{eqnarray}}
\def \eer  {\end{eqnarray}}

\begin{document}
\newcommand{\newc}{\newcommand}

\newc{\be}{\begin{equation}}
\newc{\ee}{\end{equation}}
\newc{\ba}{\begin{eqnarray}}
\newc{\ea}{\end{eqnarray}}
\newc{\bea}{\begin{eqnarray*}}
\newc{\eea}{\end{eqnarray*}}
\newc{\D}{\partial}
\newc{\ie}{{\it i.e.} }
\newc{\eg}{{\it e.g.} }
\newc{\etc}{{\it etc.} }
\newc{\etal}{{\it et al.}}
\newc{\lcdm }{$\Lambda$CDM }
\newcommand{\nn}{\nonumber}
\newc{\ra}{\rightarrow}
\newc{\lra}{\leftrightarrow}
\newc{\lsim}{\buildrel{<}\over{\sim}}
\newc{\gsim}{\buildrel{>}\over{\sim}}
\title{Six Puzzles for LCDM Cosmology}
\author{L. Perivolaropoulos}
 \affiliation{Department of Physics, University of Ioannina, Greece}
\date{\today}

\begin{abstract}
The \lcdm cosmological model is a well defined, simple and predictive model which is consistent with the majority of current cosmological observations. Despite of these successes there are specific cosmological observations which differ from the predictions of \lcdm at a level of $2\sigma$ or higher. These observations include the following: 1. Large Scale Velocity Flows (\lcdm predicts significantly smaller amplitude and scale of flows than what observations indicate), 2. Brightness of Type Ia Supernovae (SnIa) at High Redshift $z$ (\lcdm predicts fainter SnIa at High $z$), 3. Emptiness of Voids (\lcdm predicts more dwarf or irregular galaxies in voids than observed), 4. Profiles of Cluster Haloes (\lcdm predicts shallow low concentration and density profiles in contrast to observations which indicate denser high concentration cluster haloes) 5. Profiles of Galaxy Haloes (\lcdm predicts halo mass profiles with cuspy cores and low outer density while lensing and dynamical observations indicate  a central core of constant density and a flattish high dark mass density outer profile), 6. Sizable Population of Disk Galaxies (\lcdm predicts a smaller fraction of disk galaxies due to recent mergers expected to disrupt cold rotationally supported disks). Even though the origin of some of the above challenges may be astrophysical or related to dark matter properties, it should be stressed that even on galactic and cluster scales, the effects of dark energy on the equilibrium and stability of astrophysical systems are not negligible and they may play a key role in the resolution of the above puzzles. Here, I briefly review these six challenges of \lcdm and discuss the possible dark energy properties required for their resolution.
\end{abstract}
\pacs{98.80.Es,98.65.Dx,98.62.Sb}
\maketitle

\section{Introduction}

Accumulating diverse observational evidence have indicated that the universe has entered a phase of accelerating expansion. Such observations include direct geometrical probes (standard candles like SnIa \cite{SN,sn2,gold06,union08}, gamma ray bursts \cite{GRB} and standard rulers like the CMB sound horizon\cite{Percival:2007yw,Komatsu:2008hk}) and dynamical probes (growth rate of cosmological perturbations \cite{growth} probed by the redshift distortion factor or by weak lensing \cite{weak-lens}).

All these observational probes are converging towards confirming the accelerating expansion of the universe assuming the homogeneity of the universe. They have ruled out at several $\sigma$ a flat matter dominated universe and they have produced excellent fits for the simplest cosmological model predicting accelerating cosmic expansion. This model is based on the assumptions of flatness, validity of general relativity, the presence of the cosmological constant $\Lambda$ and Cold Dark Matter (\lcdm)\cite{lcdm-rev}.

From  the theoretical viewpoint the main weak points of \lcdm include \cite{lcdm-rev}:\begin{itemize} \item The Fine Tuning Problem: What is the physical mechanism that sets the value of $\Lambda$ to its observed value which is 120 orders of magnitude smaller than the physically anticipated value? \item The Coincidence Problem: Why is the energy density corresponding to the cosmological constant just starting to dominate the universe at the present cosmological time? \end{itemize} Despite of efforts to increase the complexity of \lcdm (using eg quintessence\cite{quintess} or modified gravity\cite{modgrav}) in order to address the above weak points there has been no successful alternative that addresses the above problems without replacing them with other similar ones involving fined tuned parameters. Since the theoretical weaknesses of the model have lead to no successful alternative it may be useful to identify the observational weak points of \lcdm and use these as a guide to building alternative models.

In view of the fact that \lcdm is a simple, well defined and predictive model, it is important and straightforward to test its validity using a wide range of observational probes. If some of these observational probes indicate inconsistency of \lcdm with observations then it is interesting to consider the modifications of the model required to establish consistency with observations.

Most approaches in testing the consistency of \lcdm with observations have focused on comparing \lcdm with alternative models or parameterizations on the basis of a bayesian analysis using the geometrical and dynamical probes mentioned above\cite{SN,sn2,gold06,union08,GRB,Percival:2007yw,Komatsu:2008hk,growth,weak-lens}. Due to its simplicity and acceptable quality of $\chi^2$ fit, \lcdm usually comes out as a winner in such a comparison \cite{bayes-lcdm}.

Despite of the simplicity and apparent consistency of \lcdm with most cosmological observations there are specific observational challenges for the model which have developed and persisted during the past few years. Some of these challenges involve galactic scale phenomena and it has been common wisdom that they will be resolved once astrophysical effects on these scales are better understood. Other challenges however, involve phenomena on scales larger than $\sim 10 h^{-1} Mpc$ and these may require more drastic modifications of the model in order to be resolved. Such large scale challenges of \lcdm include the observed high amplitude of large scale velocity flows on scales $\gsim 100h^{-1} Mpc$ \cite{Watkins:2008hf,Kashlinsky:2008ut,Lavaux:2008th}, the unexpected brightness of high redshift Type Ia supernovae (SnIa)\cite{Perivolaropoulos:2008yc}, the halos of massive clusters of galaxies which are more concentrated and denser than predicted by \lcdm \cite{Broadhurst:2008re} and the emptiness of voids which is unexpected in the context of \lcdm \cite{Tikhonov:2008ss,Peebles:2003pk}. On smaller (galactic) scales \lcdm is challenged by observations of constant density galactic halo cores instead of the \lcdm predicted cuspy central cores \cite{deBlok:2005qh}, the higher than expected density of outer galactic haloes \cite{Gentile:2007sb} and the sizable population of cold rotationally supported disk galaxies \cite{Bullock:2008wv}.

Since the above effects are statistically significant at $2\sigma$ level or more it is unlikely that they are all statistical fluctuations. In fact, it is possible that the resolution of the above puzzles will require more than a better understanding of astrophysical effects present on galactic scales. It may require a significant modification of the cosmological scale properties of the standard \lcdm model such as the properties of gravity, dark energy or dark matter.

The goal of the present paper is to review the above phenomena challenging the foundations of the standard \lcdm cosmological model. I will also discuss possible features of the model that may require modification in order to improve consistency with the above observations.

It should be stressed that this is not a complete list of cosmological puzzles related to the standard \lcdm cosmological model. There are other challenges related to the statistical isotropy of the CMB and the Axis of Evil \cite{aoe} (anomalous alignment of CMB multipoles in the direction $l\simeq -100^\circ$, $b=60^\circ$) which may be less related to the properties of dark matter or dark energy.  Such challenges are not discussed in the present brief review even though they may be related to the high amplitude and coherence bulk flows discussed in the next section.

\section{Challenging \lcdm}
\subsection{Large Scale Velocity Flows}
The bulk flow corresponding to the CMB dipole is closely related to the amplitude and growth rate of fluctuations
on large scales, and can be used to test cosmological models \cite{vitorio}. A number of large scale velocity surveys have been undertaken \cite{lsvf} in the past two decades and a significant amount of peculiar velocity data on a wide range of scales is currently available. The issue of comparing such sparse surveys with expectations
from cosmological models has also been investigated by several studies \cite{clsvf}.

A combined sample of peculiar velocity data has been recently used \cite{Kashlinsky:2008ut,Watkins:2008hf} to investigate the amplitude and coherence scale of the dipole bulk flow.  It was found that the dipole moment (bulk flow) of the combined  sample extends \cite{Watkins:2008hf} on scales up to $100 h^{-1}Mpc$ ($z\leq 0.03$) and perhaps up to $600h^{-1}Mpc$ ($z<0.2$ \cite{Kashlinsky:2008ut}) with amplitude larger than $400 km/sec$ \cite{Watkins:2008hf} (perhaps up to $1000km/sec$ \cite{Kashlinsky:2008ut}). The direction of the flow has been found consistently to be approximately in the direction $l \simeq 285^\circ$, $b\simeq 10^\circ$ in rough agreement with the CMB dipole ($l \simeq 276^\circ$, $b\simeq 30^\circ$). Similar results implying a large bulk low $\sim 500 km/sec$ on scales up to $100 h^{-1}Mpc$ were recently also obtained in Ref. \cite{Lavaux:2008th}. The expected $rms$ bulk flow in the context of \lcdm normalized with WMAP5 $(\omm,\sigma_8)=(0.258,0.796)$ on scales larger than $50h^{-1}Mpc$ is approximately $110 km/sec$ while the probability that a flow magnitude larger than $400km/sec$ is realized in the context of the above \lcdm normalization on scales larger than $50h^{-1} Mpc$ is less than $1\%$.

This is also demonstrated in Fig. 1 (from Ref. \cite{Watkins:2008hf}) where the $(\omm,\sigma_8)$ $\chi^2$ confidence contours obtained from the observed velocity flows (dashed lines) are superposed with the corresponding contours obtained from WMAP5 data (blue solid lines) and from WMAP5+Baryon Acoustic Oscillations+SnIa (WMAP5+BAO+SN: red dashed line).The probability of consistency of bulk flow data with \lcdm would be even lower if the data of Ref. \cite{Kashlinsky:2008ut} were considered where a flow of more than $600km/sec$ was observed on scales of $\sim 600 h^{-1}Mpc$.
\begin{figure}[!t]
\hspace{0pt}\rotatebox{0}{\resizebox{.5\textwidth}{!}{\includegraphics{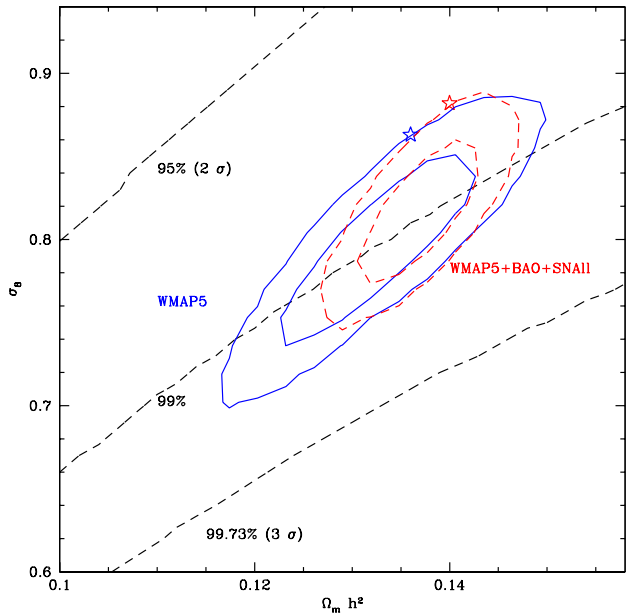}}}
\vspace{0pt}{\caption{The $(\omm,\sigma_8)$ $\chi^2$ confidence contours obtained from the observed velocity flows (dashed lines) \cite{Watkins:2008hf} are superposed with the corresponding contours obtained from WMAP5 data (blue solid lines) and from WMAP5+BAO+SN (red dashed line) (from Ref. \cite{Watkins:2008hf}).}} \label{fig1}
\end{figure}

A potential resolution of the above described conflict between the high $z$ WMAP5 normalization of \lcdm and the low $z$ normalization implied by the observed bulk flows could involve the existence of superhorizon sized non-Gaussian
and non-inflationary inhomogeneities \cite{MersiniHoughton:2008rq}, a large void at distances of order gigaparsecs \cite{GarciaBellido:2008yq}, or a redshift dependent $\sigma_8$ which changes by a factor of 2 between high $z$ and low $z$ due to an unknown physical reason. Other possibilities include a very large statistical fluctuation, a redshift dependence of Newton's constant or a redshift dependence of the dark energy equation of state parameter $w=w(z)$ leading to amplified gravity and dark energy clustering at early times ($w(z)>-1$ at $z>0.2$).

\subsection{Bright High $z$ SnIa}
As discussed in the introduction, geometrical tests of \lcdm usually involve a bayesian comparison of \lcdm with other dark energy parametrizations. This approach has not revealed so far any statistically significant weak points of the model with respect to the geometrical and dynamical probes considered.

Apart from the bayesian analysis approach, the \lcdm model can be tested by comparing the real SnIa data with Monte Carlo simulations consisting of fictitious cosmological data that would have been obtained in the context of a \lcdm cosmology. This comparison can be made on the basis of various statistics which attempt to pick up features of the data that can be reproduced with difficulty by a \lcdm cosmology\cite{Perivolaropoulos:2008yc}.
\begin{figure}[!t]
\hspace{0pt}\rotatebox{0}{\resizebox{.5\textwidth}{!}{\includegraphics{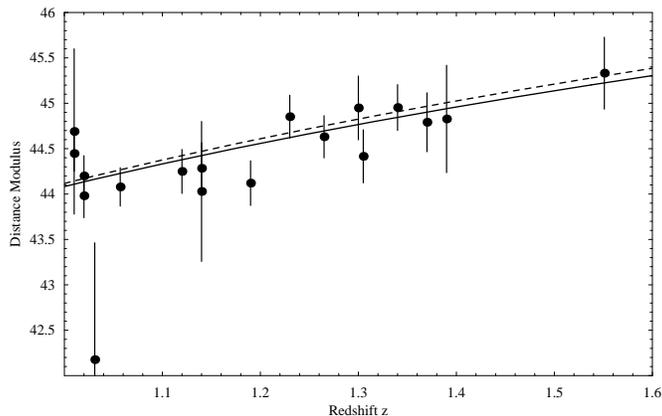}}}
\vspace{0pt}{\caption{The Union08\cite{union08} distance moduli data superposed with the best fit \lcdm model ($\omm=0.29$) dashed line and with the best fit  $(w_0,w_1)=(-1.4,2)$ model ($\omm=0.30$) continous line. Notice that at high redshifts z the distance moduli tend to be below the \lcdm  best fit while the trend is milder in the PDL crossing best fit model (from Ref. \cite{Perivolaropoulos:2008yc}.}} \label{fig2}
\end{figure}
The existence of such features is hinted by the form of the likelihood contours in various parameter planes containing parameter values corresponding to flat $\Lambda$CDM. For example, most SnIa datasets producing likelihood contours in the $\Omega_\Lambda - \Omega_m$ parameter plane have the $1\sigma$ contour barely intersect the line of flatness  $\Omega_\Lambda + \Omega_m = 1$ at the lower left side of the contour \cite{gold06,union08}. Similarly, likelihood contours based on either SnIa standard candles or standard rulers (CMB sound horizon or Baryon Acoustic Oscillations (BAO)) and constraining the parametrization \cite{Chevallier:2000qy} \be w(z)=w_0+ w_1 \frac{z}{1+z} \label{cpl} \ee systematically have the point corresponding to \lcdm $(w_0,w_1)=(-1,0)$ at the lower right edge of the $1\sigma$ contour while the best fit involves $w_0 <-1$, $w_1 >0$ \cite{gold06,union08,cross-evid,data-cross}. This feature has persisted consistently over the last decade and over different accelerating expansion probes \cite{cross-evid} (SnIa standard candles and CMB-BAO standard rulers). Even though the statistical significance of these features when viewed individually is relatively low, their persistent appearance makes it likely that there are systematic differences between the cosmological data and \lcdm predictions.
\begin{figure*}[ht]
\centering
\begin{center}
$\begin{array}{@{\hspace{-0.1in}}c@{\hspace{0.0in}}c}
\multicolumn{1}{l}{\mbox{}} &
\multicolumn{1}{l}{\mbox{}} \\ [-0.20in]
\epsfxsize=3.3in
\epsffile{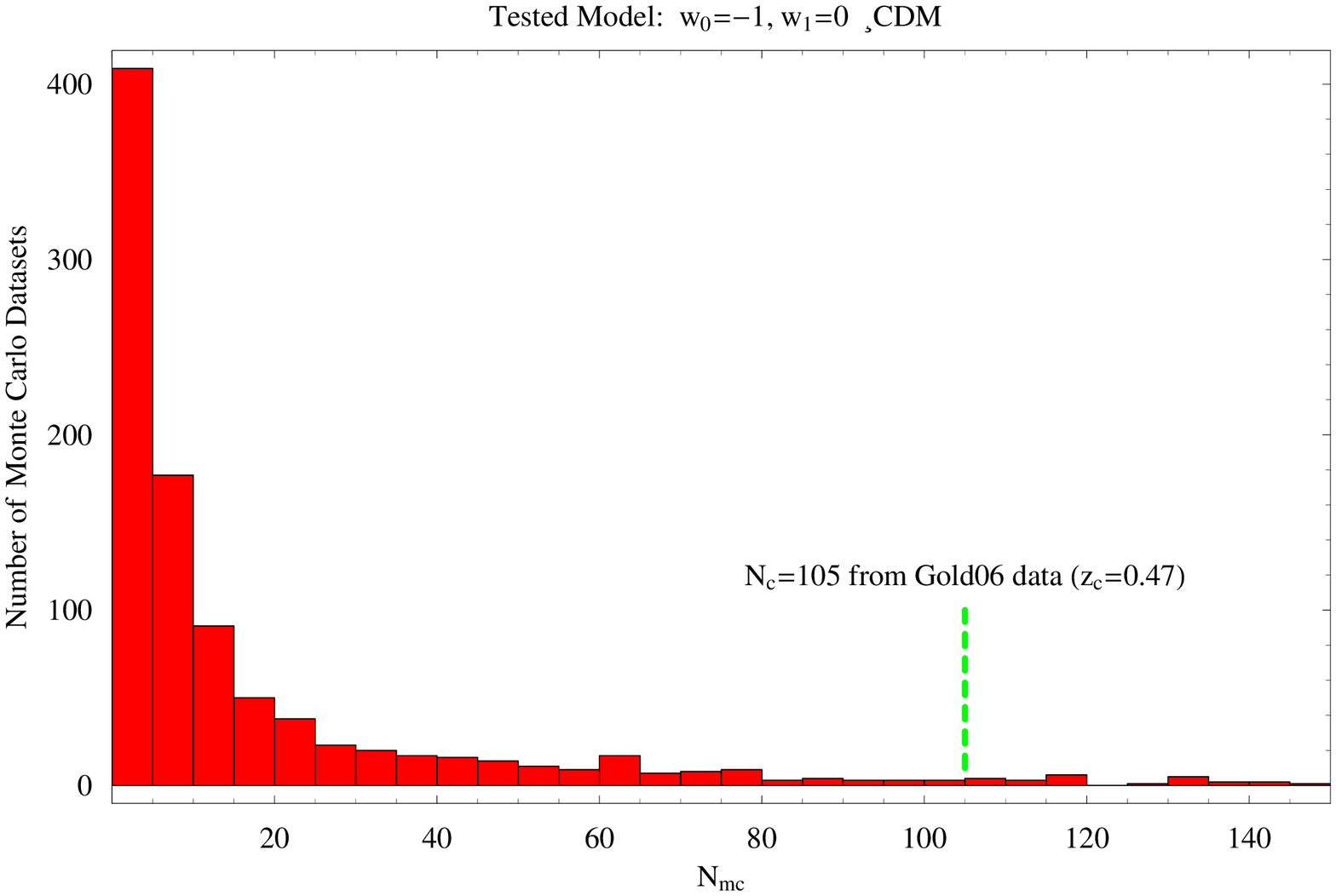} &
\epsfxsize=3.3in
\epsffile{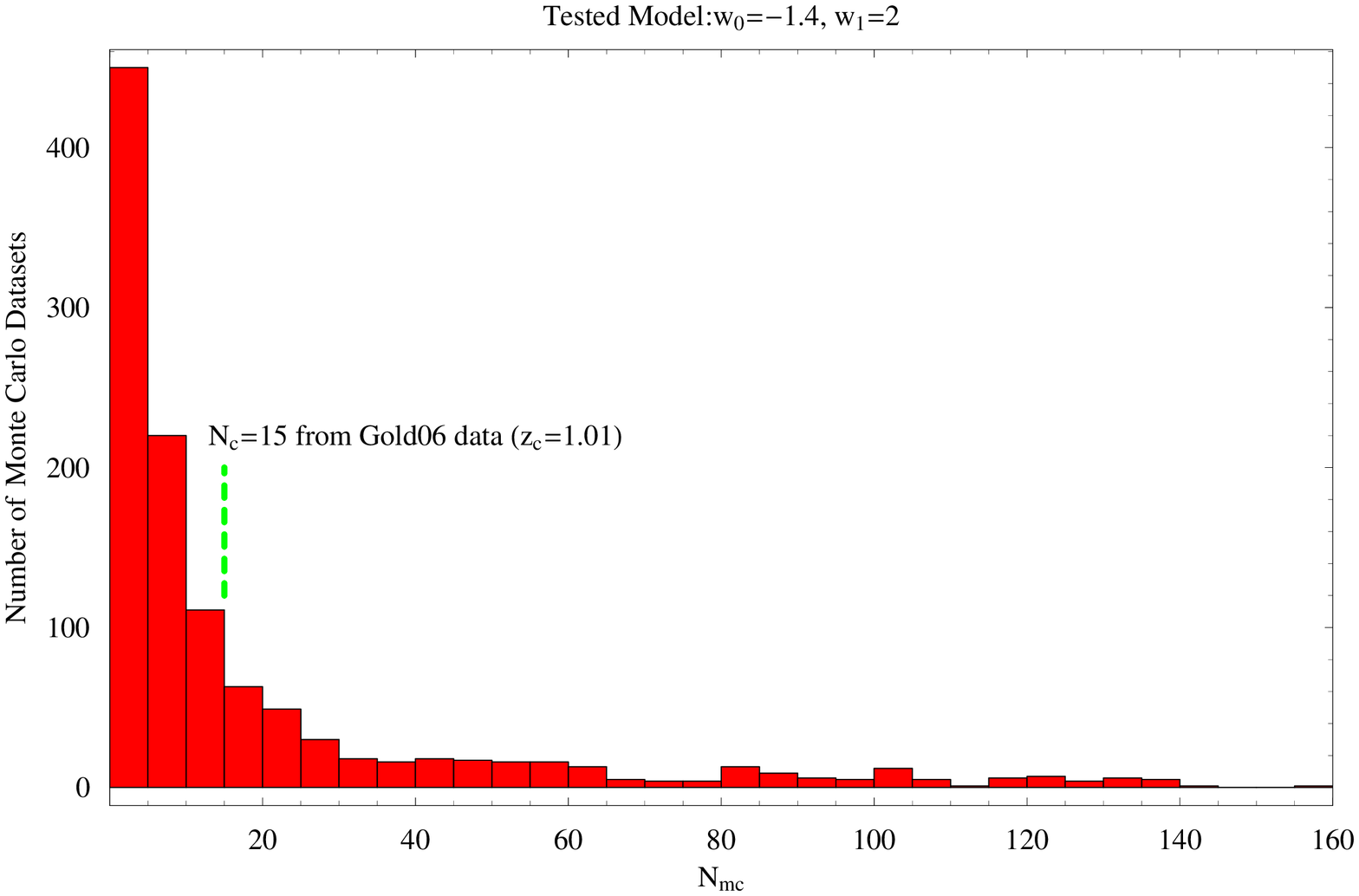} \\
\end{array}$
\end{center}
\vspace{0.0cm}
\caption{\small a: A histogram of the probability distribution of $N_{mc}$ obtained using Monte Carlo \lcdm data ($\omm=0.34$) in the context of the Gold06\cite{gold06} dataset. The thick green dashed line
corresponds to the crossing redshift $z_c$ of the real Gold06 data. b: Similar histogram for the PDL crossing model $(w_0,w_1)=(-1.4,2)$
(best fit $\omm=0.34$) instead of \lcdm. Notice that the crossing redshift $z_c$ corresponding to the real Gold06 data is a much more probable event in
the context of this cosmological model (from Ref.\cite{Perivolaropoulos:2008yc}).}
\label{fig3}
\end{figure*}

One such difference in the context of SnIa data has been recently pointed out by Kowalsky et. al. \cite{union08} where it was stated that there is `an unexpected brightness of SnIa data at $z>1$'. This feature is even directly visible by observing the SnIa distance moduli superposed with the best fit \lcdm model (dashed line in Fig. 2) where most high $z$ moduli are below the best fit \lcdm curve (obviously the reverse happens at low redshifts to achieve a good fit). Notice that this bias is smaller in the context of a parametrization that crosses the PDL $w=-1$ (continuous line in Fig. 2). \footnote{In the PDL crossing model we fix $w_0$, $w_1$ and vary $\omm$ only, in order to mimic the \lcdm number of parameters.}

This anomalous behavior of the data with respect to the \lcdm best fit may be attributed to the systematic brightness trend of high redshift SnIa with respect to the best fit \lcdm model. It is likely that this bias of the SnIa data with respect to \lcdm best fit is also responsible for the systematic mild preference (at $1\sigma$) of the SnIa data for a $w(z)$ crossing the $w=-1$ line.

In order to study quantitatively the likelihood of the existence of the above described bias in the context of a \lcdm cosmology,  we may use a statistic\cite{Perivolaropoulos:2008yc} (the Binned Normalized Differences (BND)) specially designed to pick up systematic brightness trends of the SnIa data with respect to a best fit cosmological model at high redshift. The BND statistic is based on binning the normalized differences between the SnIa distance moduli and the corresponding best fit values in the context of a specific cosmological model (eg $\Lambda CDM$). These differences are normalized by the standard errors of the observed distance moduli. We then focus on the highest redshift bin and extend its size towards lower redshifts until the Binned Normalized Difference (BND) changes sign (crosses 0) at a redshift $z_c$ (bin size $N_c$). The bin size $N_c$ of this crossing (the statistical variable) is then compared with the corresponding crossing bin size $N_{mc}$ for Monte Carlo data realizations based on the best fit model. It may be shown\cite{Perivolaropoulos:2008yc} that the crossing bin size $N_c$ obtained from the Union08 and Gold06 data with respect to the best fit \lcdm model is anomalously large compared to $N_{mc}$ of the corresponding Monte Carlo datasets obtained from the best fit \lcdm in each case. In particular, only $2.2\%$ of the Monte Carlo \lcdm datasets are consistent with the Gold06 value of $N_c$ (see Fig. 3a) while the corresponding probability for the Union08 value of $N_c$ is $5.3\%$. Thus, according to this statistic, the probability that the high redshift brightness bias of the Union08 and Gold06 datasets is realized in the context of a $(w_0,w_1)=(-1,0)$ model (\lcdm cosmology) is less than $6\%$. The corresponding realization probability in the context of a $(w_0,w_1)=(-1.4,2)$ model is more than $30\%$ for both the Union08 and the Gold06 (see Fig. 3b) datasets indicating a much better consistency for this model with respect to the BND statistic.

This result reveals a potential challenge for \lcdm cosmology and provides the motivation for obtaining additional SnIa data at high redshifts $z>1$ which may confirm or disprove the anomalous high $z$ SnIa brightness which is mainly responsible for the low probability of the high $z$ SnIa data in the context of $\Lambda CDM$.

Clearly, the unexpected high $z$ brightness of SnIa can be interpreted either as a trend towards more deceleration at high $z$ than expected in the context of \lcdm or as a statistical fluctuation or finally as a systematic effect perhaps due to a mild SnIa evolution at high $z$. However, in view of the fact that a similar mild trend for more  deceleration than expected at high $z$ is also observed in the context of standard rulers \cite{cross-evid,Percival:2007yw,data-cross}, the latter two interpretations are less likely than the first.

\subsection{The Emptiness of Voids}

Cosmological simulations performed in the context of \lcdm predict \cite{Gottloeber:2003zb} that many small dark matter haloes should reside in voids\cite{Peebles:2003pk,Peebles:2007qe,Tikhonov:2008ss}. This is consistent with observations on large scales involving giant voids defined by $10^{12}M_{\odot}$ haloes \cite{Patiri:2006gr}. Smaller voids however ($\sim 10Mpc$) look very empty. Dwarf galaxies do not show a tendency to fill these voids even though \lcdm predicts that many dwarf dark matter haloes should be in the voids. In fact as discussed in \cite{Peebles:2007qe}, if efficiency of conversion of \lcdm halos to galaxies observable in optical or HI emission were independent of environment then we would expect that about ten galaxies with $-18 < M_B < -10$ (extreme dwarfs) are in the Local Void while none is observed\cite{Karachentsev:2004dx}.

This dwarf galaxy overabundance problem of \lcdm (the `void phenomenon'\cite{Peebles:2001nv})
is also connected to the predicted number of dwarf satellites in the Local Group (the `missing satellite problem'\cite{Klypin:1999uc}): the theory predicts a factor of ten more haloes as compared with the
observed number of dwarf galaxies. For example the \lcdm model predicts
that thousands of dwarf dark matter haloes should exist in the Local
Group \cite{Klypin:1999uc,Moore:2001fc,Madau:2008fr}, while only $\sim 50$ are observed. Recent discoveries of
very low luminosity dwarfs \cite{Simon:2007dq} and careful analysis of incompleteness effects in SDSS \cite{Strigari:2007ma,Simon:2007dq} bring the theory and observations a bit closer, but the mismatch seems is still present. 

Potential resolutions of the above tension between \lcdm theory and observations involve incompleteness of observational sample, failure of many dwarf haloes to form stars\cite{Tinker:2008pr} in the context of a mass dependent bias model (Halo Occupation Distribution\cite{Seljak:2000gq})
 or peculiar properties of dark matter and/or dark energy which accelerate growth of perturbations and allow gravity to clean up voids at early times. Even though the simplest resolution of the 'void phenomenon' could involve the use of a simple bias model where galaxy
formation is driven predominantly by the mass of the host dark matter halo \cite{Tinker:2008pr}, the effectiveness of this approach at the level of dwarf galaxy luminosities is under debate\cite{Peebles:2007qe,Tikhonov:2008ss} and it is possible that an environment dependent bias may be required.

\subsection{Galaxy Halo Profiles}

The \lcdm theory predicts that dark matter
halos have a specific density distribution that follows the
well-known Navarro, Frenk, White (NFW) \cite{Navarro:1995iw,Navarro:1996gj} profile:
\be
\rho_{NFW}(R) =
\frac{\rho_s}{(R/r_s)(1 + R/r_s)^2} \label{nfw} \ee
where $r_s$ and $\rho_s$ are the characteristic radius and density of
the distribution. A useful parameter characterizing the profile is the concentration parameter $c$ defined as is $c=r_{vir}/r_s$ where $r_{vir}$ is the virial radius of the system. $r_s$ and $\rho_s$ are related to each other (e.g. \cite{Wechsler:2001cs}), so eq. (\ref{nfw}) is rather a one-parameter family of profiles.

A quite remarkable number of observations show
that NFW profiles, displaying an inner "cusp", are inconsistent
with data\cite{cuspygal}. In fact, the latter indicate profiles with a different
characteristic, a central density "core", i.e. a region where the
dark matter density remains approximately constant.

In addition to the above well-known evidence for which
in the inner regions of galaxies ($R < 2r_d$ where $r_d$ is the stellar disk radius) the dark matter haloes
show a flattish density profile, with amplitudes up to one order
of magnitude lower than the \lcdm predictions, at outer
radii ($R > 4 r_d$) the measured dark matter halo densities are found
higher than the corresponding \lcdm ones.
The dark matter halo density, known to have a
core in the internal regions, does not seem to converge to the
NFW profile at $4-6 r_d$ \cite{Gentile:2007sb}. This implies an issue for \lcdm that
should be investigated in the future, when, due to improved observational
techniques, the kinematic information will be extended
to the $~100 kpc$ scale.

A possible resolution of the puzzle of higher than expected dark matter halo density in the galactic haloes is that massive halos themselves were assembled at high redshift\cite{Peebles:2003pk}. If this is the case, modifying the properties of dark energy could play a role in shifting the epoch of galaxy formation towards earlier times. Alternatively, modified gravity theories or clustering of dark energy may also be considered as a potential resolution of this puzzle.

\subsection{Cluster Halo Profiles}

In the \lcdm context, detailed N-body simulations
have established a clear prediction that CDM-dominated cluster
halos should have relatively shallow, low-concentration
mass profiles, where the logarithmic gradient flattens continuously
toward the center with a central slope tending towards
$r^{-1}$, interior to a characteristic radius, $r_s \sim 100 - 200kpc\cdot h^{-1}$
\cite{Navarro:1995iw,Navarro:1996gj,Neto:2007vq,Bullock:1999he,Maccio':2006nu,Hennawi:2005bm,Duffy:2008pz}.

Multiply-lensed images of various clusters \cite{Broadhurst:2008re} have been used
to derive the inner mass profile \cite{Broadhurst:2004hu}, with
the outer profile determined from weak lensing \cite{Broadhurst:2004bi}. Together, the full profile has the predicted NFW
form \cite{Navarro:1996gj}, but with a surprisingly high concentration $c=\frac{r_{vir}}{r_s}$ and high density when compared
to the shallow profiles of the standard \lcdm model \cite{Broadhurst:2004bi,Umetsu:2007pq}. This result is verified  by using not only the lensing based mass profile but also the X-ray and dynamical structure in model independent analyses \cite{Lemze:2007gh}.

A potential resolution of the above discrepancy between observed cluster profiles and \lcdm predictions is that the central region of clusters collapsed, as in the case of galaxies, earlier than expected ie at $z > 1$, significantly earlier than in the standard
\lcdm, for which clusters form at $z< 0.5$.

The presence of massive clusters at high redshift ($z \sim 2$),
and the old ages of their member galaxies \cite{zirm1,Blakeslee:2003ad}, may also imply clusters collapsed at
relatively early times \cite{Mathis:2004vi}, for which
accelerated growth factors have been proposed, adopting a
generalized equation of state for dark energy \cite{Sadeh:2008br}. Such an equation of state would allow for a non-negligible dark energy density at early times. Thus, as in the case of galaxy formation, the properties of dark matter and/or dark energy could also play a significant role in the resolution of this puzzle.

\subsection{Overpopulation of Disk Galaxies}

Roughly $70\%$ of Milky-Way size dark matter halos are believed to host late-type, disk dominated
galaxies \cite{Weinmann:2005cb}. Conventional wisdom dictates that disk galaxies result from fairly
quiescent formation histories, and this has raised concerns about disk formation within the
hierarchical \lcdm cosmology \cite{Wyse:2000ym,Kormendy:2005ay}. Recent evidence for the existence of a sizeable population of
cold, rotationally supported disk galaxies at $z \sim 1.6 $ \cite{Wright:2008vf} is
particularly striking, given that the fraction of galaxies with recent mergers is expected
to be significantly higher at that time \cite{Stewart:2008ep}. High-resolution, dissipationless N-body
simulations\cite{disk1} studying the response of stellar Milky-Way type disks to such common mergers
 show that thin disks do not survive the bombardment. The remnant galaxies are roughly
three times as thick and twice as kinematically hot as the observed thin disk of the Milky Way.
However, despite of such indications a real evaluation of the severity of the problem is limited by both
theoretical and observational concerns.

The role of dark energy in the resolution of this and other astrophysical scale puzzles should not be underestimated. For example, it has been demonstrated that the effects of dark energy on the equilibrium and stability of astrophysical
structures is not negligible, and can be of relevance to describe features of astrophysical systems such as globular
clusters, galaxy clusters or even galaxies \cite{Chernin:2007cd,Iorio:2005vw,Nowakowski:2006bw,BalagueraAntolinez:2005wg}. It has recently been demonstrated that the dark energy fluid changes certain aspects of astrophysical hydrostatic equilibrium. For example, the instability of previously viable astrophysical systems when dark energy is included has been demonstrated as due to the repulsive non local dark energy force acting on the matter distribution \cite{BalagueraAntolinez:2007sf}. With the proper evolution of the dark energy equation of state, this repulsive force may also lead to a modification of the profile of the virialized structures thus addressing some of the above discussed puzzles on galactic and cluster scales.

\section{Discussion - Conclusion}

I have reviewed six of the potential observational challenges for the \lcdm cosmological model (as normalized by WMAP5) pointing out that there are such challenges on both large and small cosmological scales.  The observations conflicting the WMAP5 normalized \lcdm model at a level of $2\sigma$ or larger include the following:
\begin{itemize}
\item Large Scale Velocity Flows (\lcdm predicts significantly smaller amplitude and scale of flows than what observations indicate), \item Brightness of Type Ia Supernovae (SnIa) at High Redshift $z$ (\lcdm predicts fainter SnIa at High $z$), \item Emptiness of Voids (\lcdm predicts more dwarf or irregular galaxies in voids than observed), \item Profiles of Cluster Haloes (\lcdm predicts shallow low concentration and density profiles in contrast to observations which indicate denser high concentration cluster haloes) \item Profiles of Galaxy Haloes (\lcdm predicts halo mass profiles with cuspy cores and low outer density while lensing and dynamical observations indicate  a central core of constant density and a flattish high dark mass density outer profile),\item Sizable Population of Disk Galaxies (\lcdm predicts a smaller fraction of disk galaxies due to recent mergers expected to disrupt cold rotationally supported disks).\end{itemize}
     Even though some of the puzzles discussed here may be resolved by more complete observations or astrophysical effects, the possible requirement of more fundamental modifications of the \lcdm model remains valid.

It is interesting to attempt to identify universal features which connect these puzzles and could therefore provide a guide for their simultaneous resolution. The large scale coherent velocity flows along with the high density dark matter haloes for both galaxies and clusters seem to hint towards a more effective mechanism for structure formation at early times ($z>1$) than implied by \lcdm. This improved effectiveness could possibly be provided by a mild evolution of Newton's constant $G$ (higher $G$ at $z>0.5$) or by an evolution of the dark energy equation of state $w$ such that $w(z)>-1$ at $z\gsim 0.5$ \cite{Sadeh:2008br}. Both of these effects are expected to amplify structure formation at early times and it would be interesting to analyze quantitatively the predictions implied by the evolution of $G$ or $w$ with respect to the velocity flow and high dark matter density puzzles. The Bright High $z$ SnIa puzzle would also benefit significantly by a mild evolution of $w$ or $G$ which would imply stronger deceleration at $z>1$ than implied by \lcdm.

The improved efficiency of gravity at early times could also help emptying the voids from dark matter haloes and their corresponding galaxies thus making theoretical predictions more consistent with observations. On the other hand, the increased gravitational acceleration would also produce higher peculiar velocities that could lead to more mass inside the voids. Therefore, the predicted emptiness of voids in models with an evolving $G$ or $w$ requires a detailed study.

In conclusion, the six puzzles for \lcdm discussed in the present study provide a fertile ground for the development of both new theoretical model predictions on the corresponding observables and new observational data that would either establish or disprove these challenges for \lcdm.

\section*{Acknowledgements}
I thank Arman Shafieloo for useful comments. This work was supported by the European Research and Training Network MRTPN-CT-2006 035863-1 (UniverseNet).

\end{document}